\documentclass[12pt]{article}
\usepackage{amsmath}
\usepackage{color}
\usepackage{fancyhdr}
\usepackage{verbatim}
\usepackage{amsfonts,longtable,mathrsfs}
\usepackage{epsfig}
\usepackage{amsfonts}
\usepackage{color}
\usepackage[numbers,sort&compress]{natbib}

\def\R{\hbox{{\rm I}\kern-0.2em{\rm R}\kern0.2em}}

\def\bn{\begin{equation}}
\def\en{\end{equation}}
\def\bny{\begin{eqnarray}}
\def\eny{\end{eqnarray}}
\def\be{\begin{eqnarray*}}
\def\ee{\end{eqnarray*}}
\def\bc{\begin{center}}
\def\ec{\end{center}}

\def\({\left(}
\def\){\right  )}
\def\[{\left[}
\def\]{\right]}
\def\bc{\begin{center}}
\def\ec{\end{center}}

\newtheorem{dfn}{Definition}[section]
\newtheorem{thm}{Theorem}[section]
\newtheorem{rem}{Remark}[section]
\newtheorem{pro}{Proposition}[section]

\newtheorem{cor}{Corollary}[section]
\newtheorem{lem}{Lemma}[section]
\newtheorem{exm}{Example}[section]

\def\bn{\begin{equation}}
\def\en{\end{equation}}
\def\bny{\begin{eqnarray}}
\def\eny{\end{eqnarray}}
\def\be{\begin{eqnarray*}}
\def\ee{\end{eqnarray*}}
\def\bdn{\begin{dfn}}
\def\edn{\end{dfn}}
\def\btm{\begin{thm}}
\def\etm{\end{thm}}
\def\bpf{\begin{proof}}
\def\epf{\end{proof}}
\def\bpn{\begin{pro}}
\def\epn{\end{pro}}
\def\brk{\begin{rem}}
\def\erk{\end{rem}}
\def\bcy{\begin{cor}}
\def\ecy{\end{cor}}
\def\blm{\begin{lem}}\def\elm{\end{lem}}
\def\bex{\begin{exm}}
\def\eex{\end{exm}}

 \def\R{{\hat R}}
\begin{document}

\bc {\bf Symmetry Analysis and Conservation Laws of some third-order Difference
Equations
  }\ec
\medskip
\bc
S Mamba, MK Folly-Gbetoula and AH Kara 
\footnote{Abdul.Kara@wits.ac.za} \vspace{1cm}
\\School of Mathematics, University of the Witwatersrand, Johannesburg, South Africa.\\

\ec
\begin{abstract}
\noindent We derive a method for finding Lie Symmetries for third-order difference equations. We use these symmetries to reduce the order of the difference equations and hence obtain the solutions of some third-order difference equations. We also introduce a technique for obtaining their first integrals.
\end{abstract}
\section{Introduction} \setcounter{equation}{0}
The concept of Lie Symmetry analysis applied onto difference equations ($\Delta$E's) was studied by P. Hydon \cite{hydon} and others see \cite{Win,Mae2, P.E, Qui}. In  \cite{hydon}, the author introduced a method for obtaining symmetries and first integrals of second-order difference equations. In this paper, we extend these findings and introduce a technique for obtaining symmetries and first integrals for third-order $\Delta$E's.
\subsection{Mathematical tools to be used}
The definitions and notations used in this paper follow from the ones adopted by Hybon in \cite{hydon}. Assume we have an $Nth-order$ difference equation of the form
\begin{equation}
u_{n+N}=\omega(n, u_n, u_{n+1},...,u_{n+N-1}),
\label{eq: a}
\end{equation}
for some function $\omega$. We will assume that $\frac{\partial \omega}{\partial u_n}\neq 0$ and that the point transformations are given by
\begin{equation}
\Gamma_{\epsilon} :x\mapsto \tilde{x}(x;\epsilon),
\label{eq: b}
\end{equation}
where $x=x_i, $ $i=1,\dots,p$ are continuous variables. We then refer to $\Gamma$ as one-parameter Lie group of transformations if it satisfies the following transformations (see \cite{11})

\begin{itemize}
\item $\Gamma_0$ is the identity map if $\tilde{x}=x$ when $\epsilon=0$
\item $\Gamma_a\Gamma_b=\Gamma_{a+b}$ for every $a$ and $b$ sufficiently close to 0
\item Each $\tilde{x_i}$ can be represented as a Taylor series (in a neighbourhood of $\epsilon=0$ that is determined by $x$), and therefore
\end{itemize}

\begin{equation}
\tilde{x_i}(x:\epsilon)=x_i+\epsilon \xi _i(x)+O(\epsilon ^2), i=1,...,p.
\label{eq: c}
\end{equation}
We shall assume that the Lie point symmetries in this article take the following form
\begin{equation}
\tilde{n}=n ;\qquad
\tilde{u}_n \simeq u_n + \epsilon Q(n, u_n)
\label{eq: d}
\end{equation}
and that the analogous infinitesimal symmetry generator takes the form
\begin{equation}
X=Q(n, u_n)\partial u_n+SQ\partial u_{n+1}+S^2Q(n, u_n)\partial u_{n+2}+...+S^{N-1}Q(n, u_n)\partial u_{n+N-1},
\label{eq: e}
\end{equation}
where $S$ is defined as the shift operator, i.e.
\begin{equation}
S:n\mapsto n+1
\label{eq: f}
\end{equation}
and $Q(n, u_n)$ is referred to as the characteristic. We shall define the symmetry condition as
\begin{equation}
\tilde{u}_{n+N}=\omega(n, \tilde{u}_{n}, \tilde{u}_{n+1},...,\tilde{u}_{n+N-1})
\label{eq: g}
\end{equation}
each time \eqref{eq: a} is true. Substituting the Lie point symmetries \eqref{eq: d} into the symmetry condition \eqref{eq: g} yields linearized symmetry condition
\begin{equation}
S^{(N)}Q-X\omega=0,
\label{eq: h}
\end{equation}
whenever \eqref{eq: a} is true. We then define the conservation laws of \eqref{eq: a} as follows
\begin{equation}
(S-id)\phi (n, u_n, u_{n+1},...,u_{n+N-1})=0,
\label{eq: i}
\end{equation}
where $id$ is the identity function each time \eqref{eq: a} holds.
\section{Finding symmetries and first integrals}
\subsection{Symmetries}
Consider the following third-order difference equation
\begin{equation}
u_{n+3}=\omega(n, u_n, u_{n+1}, u_{n+2}).
\label{eq: 1}
\end{equation}
The linearized symmetry condition then amounts to
\begin{equation}
S^3Q-X\omega=0,
\label{eq: 2}
\end{equation}
where
\begin{equation}
X=Q(n, u_n)\partial u_n+SQ(n, u_n)\partial u_{n+1}+S^2Q(n, u_n)\partial u_{n+2},
\label{eq: 3}
\end{equation}

is the analogous symmetry generator. In order to obtain the characteristic $Q(n, u_n),$ we apply \eqref{eq: 2} to \eqref{eq: 1} to obtain
\begin{equation}
Q(n+3, \omega)-\frac{\partial \omega}{\partial u_n}Q(n, u_n)-\frac{\partial \omega}{\partial u_{n+1}}SQ(n, u_n)-\frac{\partial \omega}{\partial u_{n+2}}S^2Q(n, u_n),
\label{eq: 4}
\end{equation}
where $\omega$ represents the left hand side of \eqref{eq: 1}. We then differentiate \eqref{eq: 4} with respect to $u_n$, keeping $\omega$ and taking $u_{n+1}$ as a function of $u_n, u_{n+1}, u_{n+2}$ and $\omega.$ We end up with the following determining equation
\begin{equation}
\begin{split}
& \frac{\omega_{,u_n}}{\omega_{,u_{n+1}}}\big[\omega_{,u_n u_{n+1}}Q(n, u_n)+\omega_{,u_{n+1}}SQ'(n, u_n)+\omega_{,u_{n+1}u_{n+1}}SQ(n, u_n)\\
&+\omega_{,u_{n+1}u_{n+2}}S^2Q(n, u_n)\big] -\omega_{,u_n}Q'(n, u_n)-\omega_{,u_n u_n}Q(n, u_n)\\
&-\omega_{,u_n u_{n+1}}SQ(n, u_n)-\omega_{,u_n u_{n+2}}S^2Q(n, u_n)=0,
\label{eq: 5}
\end{split}
\end{equation}

where $g_{,x}$ denotes the derivative of $g$ with respect to $x.$
Note that \eqref{eq: 1} may sometimes be independent of $u_{n+1}$ or $u_{n+2},$ and so the above determining equation changes. For an example if \eqref{eq: 1} is independent of $u_{n+1},$ the resulting determining equation is
\begin{equation}
\begin{split}
& \frac{\omega_{,u_n}}{\omega_{,u_{n+2}}}\left[\omega_{,u_n u_{n+2}}Q(n, u_n)+\omega_{,u_{n+2}}S^2Q'(n, u_n)+\omega_{,u_{n+2}u_{n+2}}S^2Q(n, u_n)\right] \\
& -\omega_{,u_n}Q'(n, u_n)-\omega_{,u_n u_n}Q(n, u_n)-\omega_{,u_n u_{n+2}}S^2Q(n, u_n)=0.
\label{eq: 6}
\end{split}
\end{equation}
\subsection{Conservation Laws}
We now introduce the general case for finding first integrals of third-order $\Delta$E's. Assume that
\begin{equation}
\phi=\phi(n, u_n, u_{n+1}, u_{n+2})
\label{eq: 7}
\end{equation}
is the first integral of \eqref{eq: 1}. According to the equation
\begin{equation}
S\phi=\phi,
\label{eq: 8}
\end{equation}
we have that
\begin{equation}
\phi(n+1, u_{n+1}, u_{n+2}, \omega)=\phi(n, u_n, u_{n+1}, u_{n+2}).
\label{eq: 9}
\end{equation}
We then define
\begin{equation}
P_0 =\frac{\partial \phi}{\partial u_n},
\qquad
P_1 =\frac{\partial \phi}{\partial u_{n+1}},
\qquad
P_2 =\frac{\partial \phi}{\partial u_{n+2}}.
\label{eq: 10}
\end{equation}

It then follows that

\begin{equation}
\begin{split}
P_0 &=\left(\frac{\partial \omega}{\partial u_n}\right) \cdot S(P_2)
\\
P_1 &=S(P_0)+\left(\frac{\partial \omega}{\partial u_{n+1}}\right)\cdot S(P_2)
\\
P_2 &=S(P_1)+\left(\frac{\partial \omega}{\partial u_{n+2}}\right)\cdot S(P_2).
\label{eq: 11}
\end{split}
\end{equation}

We then join \eqref{eq: 11} to form a single equation involving only $P_2,$ which will be the determining equation for computing first integrals. The equation is as follows
\begin{equation}
S^2(\omega_{,u_n})S^3(P_2)+S(\omega_{,u_{n+1}})S^2(P_2)+\omega_{,u_{n+2}}S(P_2)-P_2=0.
\label{eq: 12}
\end{equation}
We can then solve for $P_2$ in \eqref{eq: 12}, hence solve \eqref{eq: 11} and integrate to obtain the first integrals.
\section{Applications}
We are going to use the determining equation \eqref{eq: 5} to obtain the symmetries of two third-order $\Delta$E's.
\subsection{Symmetries and exact solutions}
\subsubsection{Example 1}
Consider the third-order difference equation
\begin{equation}
u_{n+3}=\frac{a u_{n+1} u_{n+2}}{u_{n+2}+b n_n},
\label{eq: 13}
\end{equation}
where $a$ and $b$ are arbitrary constants. The determining equation \eqref{eq: 5} becomes

\begin{align}
\begin{split}
&\frac{-u_{n+1}}{bu_n+u_{n+2}}\left[-\frac{bu_{n+2}}{(bu_n+u_{n+2})^2}Q
+\frac{u_{n+2}}{bu_n+u_{n+2}}SQ'+\frac{bu_n}{(bu_n+u_{n+2})^2}S^2Q\right]\\
&+\frac{u_{n+1}u_{n+2}}{(bu_n+u_{n+2})^2}Q'-\frac{2bu_{n+1}u_{n+2}}{(u_{n+2}+bu_n)^3}Q
+\frac{u_{n+2}}{(bu_n+u_{n+2})^2}SQ\\&+\frac{u_{n+1}(bu_n-u_{n+2})}{(bu_n+u_{n+2})^3}=0.
\label{eq: 14}
\end{split}
\end{align}

Multiplying by $(bu_n+u_{n+2})^3$ throughout and differentiating twice with respect to $u_n$ we get the following differential equation
\begin{equation}
bQ^{(2)}(n, u_n)+(bu_n+u_{n+2})Q^{(3)}=0.
\label{eq: 15}
\end{equation}
Separating \eqref{eq: 15} with respect to $u_{n+2}$ we get that
\begin{equation}
Q^{(2)}=0,
\label{eq: 16}
\end{equation}
whose most general solution is
\begin{equation}
Q(n, u_n)=c_1(n)+u_nc_2(n).
\label{eq: 17}
\end{equation}
To obtain the nature of the constants $c_1,$ and $c_2,$ which are functions of $n,$ we substitute \eqref{eq: 17} back into \eqref{eq: 4} and \eqref{eq: 14}. We find that
\begin{equation}
c_1(n)=0
\qquad
c_2(n)=k_0+k_1b^{\frac{n}{2}}+(-\sqrt{b})^nk_2,
\label{eq: 18}
\end{equation}
where $k_1$ and $k_2$ are arbitrary constants. It then follows after setting $b=1$ without loss of generosity that
\begin{equation}
Q(n, u_n)=u_nk_1+(-1)^nu_nk_2.
\label{eq: 19}
\end{equation}
We therefore have that the symmetries of \eqref{eq: 13} are
\begin{equation}
\begin{split}
X_1 &=u_n\partial u_n+u_{n+1}\partial u_{n+1}+u_{n+2}\partial u_{n+2}
\\
X_2 &=u_n(-1)^n\partial u_n+u_{n+1}(-1)^{n+1}\partial u_{n+1}+u_{n+2}(-1)^{n+2}\partial u_{n+2}
\label{eq: 20}
\end{split}
\end{equation}
We note that $X_1$ is a symmetry for any value of $b$.
From $X_1$ in \eqref{eq: 20}, the resulting invariants are
\begin{equation}
\frac{du_n}{u_n}=\frac{du_{n+1}}{u_{n+1}}=\frac{du_{n+2}}{u_{n+2}}=\frac{dv_n}{0}.
\label{eq: 21}
\end{equation}
\begin{itemize}
\item
Working with the first $\left(\frac{du_n}{u_n} \right)$ and second $\left(\frac{du_{n+1}}{u_{n+1}}\right)$ invariants, first $\left(\frac{du_n}{u_n}\right)$ and third $\left(\frac{du_{n+2}}{u_{n+2}}\right)$ invariants we get that
\begin{equation}
K_1=\frac{u_n}{u_{n+1}} \qquad
K_2=\frac{u_n}{u_{n+2}} \qquad
v_n=f(K_1, K_2),
\label{eq: 22}
\end{equation}
where $K_1$ and $K_2$ are constants.
\begin{itemize}
\item[-]
We choose $f(K_1, K_2)=K_1\cdot K_2 $, therefore
\begin{equation}
v_n=\frac{u_n}{u_{n+2}},
\label{eq: 23}
\end{equation}
hence
\begin{equation}
v_{n+1}=\frac{1+bv_n}{a}.
\label{eq: 24}
\end{equation}
Solving \eqref{eq: 24} results in
\begin{subequations}
\begin{eqnarray}
v_n=&c_1\left(\frac{b}{a}\right)^{n-1}-\frac{1-\left(\frac{b}{a}\right)^n}{b-a}\qquad \text{ if } a\neq b \label{eq: 25}
\\
v_n=&c_1+ \frac{n}{b} \qquad \qquad \qquad \quad \text{ if } a=b. \label{eq:26}
\end{eqnarray}
\end{subequations}
We have reduced a third-order difference equation \eqref{eq: 1} into a second-order difference equation \eqref{eq: 25}.\\
\underline{\textit{Case a=b}:}
Combining (\ref{eq: 23}) and (\ref{eq:26}), we have
\begin{equation}\label{eq: 35}
u_{n+2}= \frac{1}{v_n}u_n=\frac{a}{ac_1+n}u_n.
\end{equation}
Therefore, the solution to (\ref{eq: 13}), when $a=b$, is given by
\begin{equation}
u_n=\frac{2^{1-\frac{n}{2}}b^{\frac{n}{2}-1}(c_1+c_2(-1)^n)}{d(\frac{bc_1}{2}+1)_{\frac{n}{2}-1}}, \quad c_2\in Z,
\end{equation}
where $v_{k_1}$ in given in (\ref{eq:26}) and $(a)_n$ is the Pochhammer symbol (rising factorial.)
\\
\underline{\textit{Case $a\neq b$}:}
 Combining (\ref{eq: 23}) and (\ref{eq: 25}), we have
\begin{equation}\label{eq: 35}
u_{n+2}= \frac{1}{v_n}u_n= \frac{1}{c_1\left(\frac{b}{a}\right)^{n-1}-\frac{1-\left(\frac{b}{a}\right)^n}{b-a}}u_n.
\end{equation}
Therefore, the solution to (\ref{eq: 13}) is given by
\begin{align}
\begin{split}
u_n=& c_2 \prod_{k_3=1}^{n-1}\exp \left((-1)^{k_3+1} \sum_{k_1=0}^{k_3-1}\left((-1)^{-k_1+1}\ln v_{k_1}\right)\right) +\\&%
 c_3\left(\sum_{k_4=0}^{n-1}\frac{ \prod_{k_2=1}^{k_4-1}-\exp\left((-1)^{k_2} \sum_{k_1=0}^{k_2}(-1)^{-k_1+1}\ln v_{k_1}\right)}{\prod_{k_3=1}^{k_4}\exp\left((-1)^{k_3+1}\sum_{k_1=0}^{k_3-1}(-1)^{-k_1+1}\ln v_{k_1}\right)}\right)\\&\prod_{k_3=1}^{n-1}\exp \left((-1)^{k_3+1} \sum_{k_1=0}^{k_3-1}\left((-1)^{-k_1+1}\ln v_{k_1}\right)\right),
 \end{split}
\end{align}
where $v_{k_1}$ in given in (\ref{eq: 25}).
\item[-]
Suppose that we choose $f(K_1, K_2)=K_1,$ following the same method above, we end up with $v_n=\frac{u_n}{u_{n+1}}$, i.e.,
\begin{equation}
v_{n+2}=\frac{1}{av_{n+1}}+\frac{bv_n}{a}=\frac{1+bv_nv_{n+2}}{av_{n+1}}.
\label{eq: 26}
\end{equation}
One can solve for $v_n$ from \eqref{eq: 26} and hence solve for $u_n.$
\end{itemize}
\item
Working with the first  $\left(\frac{du_n}{u_n} \right)$ and third invariant $\left(\frac{du_{n+2}}{u_{n+2}}\right),$ second invariant $\left(\frac{du_{n+1}}{u_{n+1}}\right)$ and third invariant $\left(\frac{du_{n+2}}{u_{n+2}}\right),$ we get that
\begin{equation}
K_1=\frac{u_n}{u_{n+2}} \qquad
K_2=\frac{u_{n+1}}{u_{n+2}} \qquad
v_n=f(K_1, K_2).
\label{eq: 27}
\end{equation}
\begin{itemize}
\item[-]
Choosing $f(K_1, K_2)=K_2/K_1 $ we get that
\begin{equation}\label{eq:41}
v_n=\frac{u_{n+1}}{u_n}
\end{equation}
 and therefore
\begin{equation}
v_{n+2}=\frac{av_n}{b+v_nv_{n+1}}=\frac{(a/b)v_n}{1+(1/b)v_nv_{n+1}}.
\label{eq: 28}
\end{equation}
The solution of (\eqref{eq: 28}) is given in \cite{MF}( put $A=a/b$ and B=1/b) by
\begin{equation}\label{eq:42}
\begin{split}
v_n=&\exp\Bigg[ (-1)^{n-1}\ln\left|{v_1}\right|+\\&(-1)^{n-1}\sum_{k_1=1}^{n-1}(-1)^{-k_1} \ln\left|\frac{v_0v_1{a/b}^{k_1}}{1+(1/b)v_0v_1(\sum_{i=0}^{k_1-1}{a/b}^i)}\right|   \Bigg]
\end{split}
\end{equation}
with $-1/((1/b)v_0v_1) \notin \{1,1+(a/b),\dots, \sum_{i=0}^{n-2} (a/b)^{i}\}$.
Combining (\ref{eq:41}) and (\ref{eq:42}) we get that
\begin{equation}\label{eq:43}
u_n=c_4 \prod_{k_1=1}^{n-1}v_{k_1},
\end{equation}
where $v_n$ is given in (\ref{eq:42}), is also a solution to (\ref{eq: 13}).
\end{itemize}
\end{itemize}
\subsubsection{Example 2}
We now shift our attention to focus on a second third-order difference equation
\begin{equation}
u_{n+3}=a(n)u_n+c(n)u_{n+2},
\label{29}
\end{equation}
where $a$ and $c$ are arbitrary functions of $n$. We note that \eqref{29} is independent of $u_{n+1},$ so we use determining equation \eqref{eq: 6}. One can easily establish that the solution to the determining equation is
\begin{equation}
Q(n, u_n)=k_1u_n+k_2(n),
\label{30}
\end{equation}
where $k_1$ is a constant and $k_2$ is a function of $n.$ Substituting \eqref{29} into \eqref{eq: 2} we get that
\begin{equation}
k_2(n+3)=a(n)k_2(n)+c(n)k_2(n+2),
\label{31}
\end{equation}
where we notice that $k_2(n)$ satisfies the original equation \eqref{29}. Assuming that $U_1,$ $U_2$ and $U_3$ are solutions of \eqref{31}, the resulting symmetries are
\begin{equation}
X_0=u_n\partial u_n \qquad
X_1=U_1\partial u_n \qquad
X_2=U_2\partial u_n \qquad
X_3=U_3\partial u_n.
\label{32}
\end{equation}
The characteristic equation, using $X_0$, is given by
\begin{equation}
\frac{du_n}{u_n}=\frac{du_{n+1}}{u_{n+1}}=\frac{du_{n+2}}{u_{n+2}}=\frac{dV_n}{0}.
\end{equation}
The invariant $V_n$ is then given by
$$V_n= g\left(\frac{u_{n+1}}{u_{n}}, \frac{u_{n+2}}{u_{n+1}} \right).$$
Assume $g(x,y)=x$, we have
\begin{equation}\label{eq: 51}
V_n=\frac{u_{n+1}}{u_n}
\end{equation}
and this implies that
\begin{equation}\label{eq: 53}
V_{n+2}=\frac{a(n)}{V_nV_{n+1}}+c(n).
\end{equation}
On the other hand, by assuming that $g(x,y)=1/x$, we have
\begin{equation}
V_n=\frac{u_{n}}{u_{n+1}}
\end{equation}
and this implies that
\begin{equation}
V_{n+2}=\frac{1}{c(n)+a(n)V_nV_{n+1}}.
\end{equation}
With these change of variables we have obtained two different reduced forms (in terms of order), given by (\eqref{eq: 51}) and (\eqref{eq: 53}),  of (\ref{29}).
\subsection{Conservation Laws}
\subsubsection{Example 1}
Consider the equation \eqref{29}, assume that one required to obtain the first integrals. By letting $P_2=P_2(n, u_n),$ one can find that the solution to the determining equation \eqref{eq: 12} is
\begin{equation}
P_2=B(n).
\label{eq: 33}
\end{equation}
Substituting \eqref{eq: 33}, we find that $B(n)$ satisfies the equation
\begin{equation}
B(n)=a(n+2)B(n+3)+c(n)B(n+1).
\label{eq: 34}
\end{equation}
We can then shown that 
\begin{align}
P_0=&a(n)B_i(n+1),\\  P_1=&a(n+1)B_i(n+2),\\ P_2=& B_i(n),
\end{align}
where $B_i's,$ $i=1,...3$ are solutions to \eqref{eq: 34}. We can therefore deduce that the first integrals of such equations \eqref{29} are
\begin{align}
\Phi_i&=P_0u_n+P_1u_{n+1}+P_2 u_{n+2}+G_i(n)\qquad \qquad \\
&= a(n)B_i(n+1)u_n+a(n+1)B_i(n+2)u_{n+1}+B_i(n)u_{n+2}+G_i(n),\label{eq: 35}
\end{align}
for some functions $G_i, i=1,2,3$, to be determined. By substituting \eqref{eq: 35} into \eqref{eq: 8}, we were able to find that $G_i(n+1)=G_i(n)=K,$ where $K$ is a constant. For simplicity, we shall assume that $K$ is zero. It has to be noted that now that we know the symmetries and first integrals of (\ref{29}), we can create new first integrals. Below are new first integrals of (\ref{29}):
\begin{align}
\Phi_{3i}=& X_1 \Phi_i =U_1P_0 + P_1 SU_1 +(S^2U_1)P_2\\
\Phi_{4i}=& X_2 \Phi_i =U_2P_0 + P_1 SU_2 +(S^2U_2)P_2\\
\Phi_{5i}=& X_3 \Phi_i =U_3P_0 + P_1 SU_3 +(S^2U_3)P_2.
\end{align}
Also, from
\begin{eqnarray}
\left( S-id \right)\Phi_ i =(SP_2) \left[ u_{n+3}-c(n)u_{n+2}-a(n)u_n \right],
\end{eqnarray}
we obtain new integrals
\begin{align}
\Phi_{6i}=& SP_2=B_i(n+1)=\frac{P_0}{a(i)}.
\end{align}
\subsubsection{Example 2}
Now consider the equation \eqref{eq: 13}, i.e.,
$$u_{n+3}=\frac{a u_{n+1} u_{n+2}}{u_{n+2}+b u_n}.$$ Assuming one required to compute the first integrals, one would hence use \eqref{eq: 12} to come up with the equation
\begin{equation}
\frac{-abu_{n+3}u_{n+4}}{(u_{n+4}+bu_{n+2})^2}S^3P_2+\frac{au_{n+3}}{u_{n+3}
+bu_{n+1}}S^2P_2+\frac{abu_nu_{n+1}}{(u_{n+2}+bu_n)^2}SP_2-P_2=0.
\label{eq: 36}
\end{equation}
Due to the fact that in the above equation, $P_2$ takes up different arguments, we first differentiate with respect to $u_n$ keeping $\omega=u_{n+3}$ fixed and by assuming $u_{n+2}$ is a function of $u_n$ and $\omega.$ We differentiate the resulting equation with respect to $u_n$ six times and we then split it with respect to $u_{n+1}$ and $u_{n+2}$. to get to equation
\begin{equation}
P_2^{(6)}(n, u_n)=0.
\label{eq: 37}
\end{equation}
After solving the differential equation \eqref{eq: 37} and substituting the result back into \eqref{eq: 36} we found that $P_2=0,$ hence $P_1=0$ and $P_0=0.$ This implies that the first integral of \eqref{eq: 13} is of a trivial nature and it can be written in the form
\begin{equation}
\phi=K,
\end{equation}
where $K$ is a constant.\par
\section{Conclusion}
We have explored a way of obtaining symmetries and first integrals of third-order difference equations. We have also demonstrated one can use the symmetries to reduce the order of the difference equations. What we have to note is that when reducing the order, we used only one of the symmetries, i.e., $u_n$ whilst we could have also used 
another symmetry which could have also worked the same way but yielded a different result.


\newpage
\begin{thebibliography}{22}
\bibitem{hydon} P. E. "Symmetries and first integrals of ordinary difference equations." Proceedings of the Royal Society of London A: Mathematical, Physical and Engineering Sciences. Vol. 456. No. 2004. The Royal Society, 2000.
\bibitem{Win} D. Levi, L. Vinet and P. Winternitz, Lie group formalism for difference equa-
tions, J. Phys. A: Math. Gen. 30 (1997) 633-649.
\bibitem{Mae2} S. Maeda. The similarity method for difference equations, IMA J. Appl. Math.38 (1987), 129-134.
\bibitem{P.E} Grant, Timothy J., and Peter E. Hydon. "Characteristics of Conservation Laws for Difference Equations." Foundations of computational mathematics 13.4 (2013): 667-692.
\bibitem {Qui} Quispel, G. R. W., and R. Sahadevan. "Lie symmetries and the integration of difference equations." Physics Letters A 184.1 (1993): 64-70.
equations, \textit{Journal of difference equations and applications}, \textbf{2016}
\bibitem{11} P. E. Hydon, Difference Equations by Differential Equation Methods, \textit{Cambridge University Press} (2014).
\bibitem{MF} MK Folly-Gbetoula. Symmetry, reductions and exact solutions of the difference equation $ u_ {n+ 2}= au_n/(1+ bu_nu_ {n+ 1}) $. preprint arXiv:1608.01878 (2016).
\end{thebibliography}
\end{document}